\documentclass[letterpaper, 10 pt, conference,onecolumn]{ieeeconf}  
\usepackage[utf8]{inputenc}
\usepackage{amsmath}
\usepackage{amsfonts}
\usepackage{amssymb}
\usepackage{mathrsfs}
\usepackage{paralist}
\usepackage{psfrag}
\usepackage{pstool}
\let\labelindent\relax
\usepackage{enumitem}
\usepackage{mathtools}
\mathtoolsset{showonlyrefs=true,showmanualtags=true}
\usepackage{color}
\usepackage{hyperref}
\usepackage{xstring}
\usepackage{verbatim}
\usepackage{siunitx}
\usepackage{cite}

\usepackage{stmaryrd}
\usepackage{scalerel}

\usepackage{stfloats}

\newcommand{\f}[1][]{f_{#1}}

\renewcommand{\v}[1][]{v_{#1}}

\let\oldalpha\alpha
\renewcommand{\alpha}[1][]{\oldalpha_{#1}}
\let\oldsigma\sigma
\renewcommand{\sigma}[1][]{\oldsigma_{#1}}
\let\oldvarsigma\varsigma
\renewcommand{\varsigma}[1][]{\oldvarsigma_{#1}}
\let\oldkappa\kappa
\renewcommand{\kappa}[1][]{\oldkappa_{#1}}
\let\oldtheta\theta
\renewcommand{\theta}[1][]{\oldtheta_{#1}}
\let\oldbeta\beta
\renewcommand{\beta}[1][]{\oldbeta_{#1}}
\let\oldgamma\gamma
\renewcommand{\gamma}[1][]{\oldgamma_{#1}}
\let\oldOmega\Omega
\renewcommand{\Omega}[1][]{\oldOmega_{#1}}
\let\oldomega\omega
\renewcommand{\omega}[1][]{\oldomega_{#1}}
\let\oldmu\mu
\renewcommand{\mu}[1][]{\oldmu_{#1}}
\let\oldnu\nu
\renewcommand{\nu}[1][]{\oldnu_{#1}}
\let\oldell\ell
\renewcommand{\ell}[1][]{\oldell_{#1}}
\let\oldeta\eta
\renewcommand{\eta}[1][]{\oldeta_{#1}}
\let\olddelta\delta
\renewcommand{\delta}[1][]{\olddelta_{#1}}
\let\oldlambda\lambda
\renewcommand{\lambda}[1][]{\oldlambda_{#1}}
\let\oldepsilon\epsilon
\renewcommand{\epsilon}[1][]{\oldepsilon_{#1}}
\let\oldvarepsilon\varepsilon
\renewcommand{\varepsilon}[1][]{\oldvarepsilon_{#1}}
\let\oldrho\rho
\renewcommand{\rho}[1][]{\oldrho_{#1}}
\let\oldzeta\zeta
\renewcommand{\zeta}[1][]{\oldzeta_{#1}}
\let\oldxi\xi
\renewcommand{\xi}[1][]{\oldxi_{#1}}
\let\oldXi\Xi
\renewcommand{\Xi}[1][]{\oldXi_{#1}}
\let\oldphi\phi
\renewcommand{\phi}[1][]{\oldphi_{#1}}
\let\oldpsi\psi
\renewcommand{\psi}[1][]{\oldpsi_{#1}}
\let\oldpi\pi
\renewcommand{\pi}[1][]{\oldpi_{#1}}
\let\oldPi\Pi
\renewcommand{\Pi}[1][]{\oldPi_{#1}}
\let\oldchi\chi
\renewcommand{\chi}[1][]{\oldchi_{#1}}
\let\oldvarphi\varphi
\renewcommand{\varphi}[1][]{\oldvarphi_{#1}}
\let\oldupsilon\upsilon
\renewcommand{\upsilon}[1][]{\oldupsilon_{#1}}

\definecolor{myblue}{rgb}{0.21, 0.52, 0.95}

\renewcommand{\hat}[1]{\widehat{#1}}
\newcommand{\lbar}[1]{\underline{#1}}
\renewcommand{\bar}[1]{\overline{#1}}
\newcommand{\pmtx}[1]{\begin{pmatrix}#1\end{pmatrix}}

\newcommand{\tto}{\rightrightarrows}

\newtheorem{theorem}{Theorem}
\newtheorem{proposition}{Proposition}
\newtheorem{corollary}{Corollary}
\newtheorem{lemma}{Lemma}
\newtheorem{assumption}{Assumption}
\newtheorem{remark}{Remark}
\newtheorem{definition}{Definition}
\newtheorem{example}{Example}

\newcommand{\ceq}{:=}
\newcommand{\ceqinv}{=:}

\newcommand{\x}{\times}
\newcommand{\inner}[2]{\langle #1,#2\rangle}
\newcommand{\norm}[2][\mbox{}]{| #2|_{#1}}

\DeclareMathOperator{\id}{id}

\newcommand{\Kinf}{\mathcal{K}_\infty}

\newcommand{\Rnneg}{\mathbb{R}_{\geq 0}}

\newcommand{\reals}[1][]{\mathbb{R}^{#1}}
\newcommand{\naturals}{\mathbb{N}}

\newcommand{\minus}{\backslash}

\DeclareMathOperator{\cl}{cl}
\DeclareMathOperator{\interior}{int}

\newcommand{\tp}{^\top}
\newcommand{\inv}{^{-1}}

\newcommand{\eigmax}{\bar{\lambda}}

\newcommand{\pd}[1]{%
	\IfEqCase{#1}{%
		{0}{p_d}%
	}[p_d^{(#1)}]
}

\newcommand{\pl}{^+}
\DeclareMathOperator{\dom}{dom}

\DeclareMathOperator{\rge}{rge}

\newcommand{\so}[1]{\mathfrak{so}(#1)}
\newcommand{\mf}[1]{\mathfrak{#1}}
\newcommand{\sk}{S}
\newcommand{\sphere}[1]{\mathbb{S}^{#1}}

\newcommand{\Amc}[1][]{\mathcal{A}_{#1}}

\newcommand{\Cmc}[1][]{\mathcal{C}_{#1}}
\newcommand{\Dmc}[1][]{\mathcal{D}_{#1}}
\newcommand{\Emc}[1][]{\mathcal{E}_{#1}}
\newcommand{\Fmc}[1][]{\mathcal{F}_{#1}}
\newcommand{\Gmc}[1][]{\mathcal{G}_{#1}}
\newcommand{\Hmc}[1][]{\mathcal{H}_{#1}}
\newcommand{\Lmc}[1][]{\mathcal{L}_{#1}}
\newcommand{\Imc}[1][]{\mathcal{I}_{#1}}
\newcommand{\Jmc}[1][]{\mathcal{J}_{#1}}

\newcommand{\Mmc}[1][]{\mathcal{M}_{#1}}

\newcommand{\Smc}[1][]{\mathcal{S}_{#1}}
\newcommand{\Tmc}[1][]{\mathcal{T}_{#1}}
\newcommand{\Qmc}[1][]{\mathcal{Q}_{#1}}
\newcommand{\Umc}[1][]{\mathcal{U}_{#1}}
\newcommand{\Wmc}[1][]{\mathcal{W}_{#1}}
\newcommand{\Xmc}[1][]{\mathcal{X}_{#1}}

\newcommand{\qed}{\hfill\square}
\IEEEoverridecommandlockouts                              

\title{\LARGE \bf
Event-Triggered Synergistic Controllers with Dwell-Time Transmission
}

\author{Xuanzhi Zhu, Pedro Casau and Carlos Silvestre
	\thanks{X. Zhu is with is with the Department of Electrical and Computer Engineering, Faculty of Science and Technology, University of Macau, Macau, China. E-mail address: {\tt\small xuanzhi.zhu@tecnico.ulisboa.pt}.
		P. Casau is with the Instituto de Telecomunica\c{c}\~{o}es and Department of Electronics, Telecommunications and Informatics, University of Aveiro, Portugal. E-mail address: {\tt\small pcasau@ua.pt}.
		C. Silvestre is with the Department of Electrical and Computer Engineering, Faculty of Science and Technology, University of Macau, Macau, China, and is on leave from the Instituto Superior T\'{e}cnico, Universidade de Lisboa, Lisboa, Portugal. E-mail address: {\tt\small csilvestre@um.edu.mo}}%
}

\begin{document}

\maketitle

\begin{abstract}
We propose novel event-triggered synergistic controllers for nonlinear continuous-time plants by incorporating event-triggered control into stabilizing synergistic controllers.
We highlight that a naive application of common event-triggering conditions may not ensure dwell-time transmission due to the joint jumping dynamics of the closed-loop system.
Under mild conditions, we develop a suite of event-triggered synergistic controllers that guarantee 
{both dwell-time transmission and global asymptotic stability.
Through numerical simulations, we demonstrate the effectiveness of our controller applied to the problem of rigid body attitude stabilization.
}

\end{abstract}

\section{Introduction}
\subsection{Background and Motivations}

Combination of {event-triggered control (cf.~\cite{Postoyan2015}) and synergistic control (cf.~\cite{Casau2024})} is rare but is attractive in practice, particularly within the field of networked control of robotic systems (cf.~\cite{Zhu2021,Zhang2022}). A properly designed event-triggered synergistic controller {must maintain} both the energy-saving feature and the global stabilization property.
This motivates us to devise a systematic way to incorporate event-triggered control into synergistic controllers, based on an \emph{emulation-like} approach as described in~\cite{Postoyan2015}.
This means, we first design stabilizing synergistic controllers by ignoring computation/communication constraints, and then derive appropriate event-triggered controllers by taking these constraints into account.

We consider the case where telecommunication consumes more power than computing so that event-triggered control is used to reduce transmissions on the communication channels.
In order to comply with the hardware specifications, it is preferable to design event-triggered controllers such that the minimum inter-transmission time is higher than the hardware's sampling time.
{However, poorly designed event-triggered controllers can lead to Zeno solutions (cf.~\cite{Casau2022}).}
When event-triggered control is implemented on hybrid controllers (like synergistic controllers), this implies that finite number of transmissions occur on each bounded time interval, but there is no guarantee of a positive lower bound of the inter-transmission times.
Indeed, the works on event-triggered implementation of synergistic controllers do not rule out the possibility of two consecutive transmissions occurring at the same continuous time (cf.~\cite{Zhu2021,Zhang2022}).
This motivates us to design event-triggered synergistic controllers that induce a positive lower bound of the inter-transmission times, which we call \emph{dwell-time transmission}. It is worth noting that this property, though closely related to the dwell-time notions in the literature {(cf.~\cite{Goebel2012,Bernard2022,Scheres2024})}, pertains to transmission events only by ignoring the synergistic switches. 

\subsection{Challenges and Approaches}\label{subsec:chaApp}
For event-triggered synergistic controllers, event-triggered control causes instantaneous updating of certain control variables, while synergistic control requires instantaneous switching of a logic variable.
These two types of jumps present two major challenges, which we address as follows.

The first challenge is to properly characterize the inter-transmission time.
Among the jumps associated with a solution, we are only interested in those that reflect transmissions.
A natural way to do so is to first omit the jumps from the synergistic controller, then reconstruct a proper hybrid arc, and finally study the elapsed time between jumps of the hybrid arc.
This observation, however, is rarely discussed in the literature.
Fortunately, a parallel can be drawn between the procedures above and the reparametrization of hybrid arcs described in~\cite[Definition~3]{Bernard2020}.
To be exact, while the latter \emph{attaches} jumps to hybrid arcs, our approach \emph{detaches} jumps from hybrid arcs.
In this pursuit, we propose a \emph{deparametrization} of solutions to address the first challenge.

The second challenge is to guarantee dwell-time transmission, which does not necessarily equal the elapsed continuous time between two consecutive jumps.
As a result, the well-known geometric-based condition $\Gmc(\Dmc)\cap\Dmc=\emptyset$ in~\cite[Proposition~2.34]{Ricardo2021} is not sufficient for dwell-time transmission. It is worth noting that, when neither the plant nor the underlying controller has jump dynamics, the condition above becomes sufficient and necessary (cf.~\cite{Chai2020}) and even uniformity of the dwell-time with respect to the initial state can be achieved (cf.~\cite{Postoyan2015}).
A natural way to relax the aforementioned condition is assuming $\Gmc^k(\Dmc)\cap\Dmc=\emptyset$ for some $k\in\naturals$, similar to~\cite[Definition~1]{Biemond2013} that handles hybrid inputs. A slightly different form is also found in~\cite[Definition~2]{Zhu2021} that proves the absence of Zeno solutions.
Meanwhile, we reexamine the pathological case of two consecutive transmissions occurring at the same continuous time (cf.~\cite{Zhu2021,Zhang2022}). The crux of this problem lies in the permission of a second jump when the state is in the jump set of the event-triggered controller.
The discussions above lay the foundation for the approach we take to tackle the second challenge, namely by imposing geometric-based conditions that forbid consecutive jumping back to the jump set of the event-triggered controller.

\subsection{Contributions}
1) We ensure dwell-time for \emph{deparametrized} hybrid arcs of solutions to a class of hybrid systems with joint jumping dynamics. It is straightforward to apply this general result to guaranteeing dwell-time transmission for event-triggered synergistic controllers;
2) We propose a suite of novel event-triggered synergistic controllers with {both dwell-time transmission guarantee and global asymptotic stabilization capability}, applying them to the practical problem of rigid body attitude stabilization.
\subsection*{Notations}Let $\naturals$, $\reals$, $\Rnneg$, $\reals[n]$, and $\reals[n\x m]$ denote the natural numbers, the real numbers, the non-negative real numbers, the $n$-dimensional Euclidean space, and the space of $n\x m$ real matrices, respectively.
Let $0,I$ denote the zero matrix and the identity matrix with appropriate dimensions.
In $\reals[n]$, the inner product is defined by $\inner{v}{w} \ceq v\tp w$ for each $v,w\in\reals[n]$, the Euclidean norm is defined by $\norm{v}=\sqrt{\inner{v}{v}}$ for each $v\in\reals[n]$, and the distance from $\v\in\reals[n]$ to $\Amc\subset\reals[n]$ is defined by $\norm[\Amc]{v}\ceq\inf_{w\in\Amc}\norm{w-v}$.
The length of an interval $\Imc\subset\reals\cup\{\pm\infty\}$ is defined by $|\Imc|\ceq\sup\Imc-\inf\Imc$.
The closure, interior, and cardinality of $\Amc\subset\reals[n]$ are denoted by $\cl(\Amc)$, $\interior(\Amc)$, and $\#(\Amc)$, respectively.
Let $\sphere{n}\ceq\{v\in\reals[n+1]:\norm{v}= 1\}$ denote the $n$-dimensional sphere.
Given $X=X\tp\in\reals[n\x n]$, let $\eigmax(X)$ denote its maximum eigenvalue.
Let $\Kinf$ denote the class of strictly increasing continuous functions $\alpha:\Rnneg\to\Rnneg$ satisfying that $\alpha(0)=0$ and $\lim_{s\to+\infty}\alpha(s)=+\infty$.
The operator $\id:\Amc\to\Amc$ with $\Amc\subset\reals[n]$ is defined by $\id(v)\ceq v$.
The tangent cone to $\Amc\subset\reals[n]$ at $v\in\reals[n]$, denoted as $\Tmc[\Amc](v)$, contains all vectors $w\in\reals[n]$ for which there exist sequences $v_i\in\Amc$, $\tau_i>0$ with $v_i\to v$, $\tau_i\searrow 0$, and $w=\lim_{i\to\infty}(v_i-v)/\tau_i$, see~\cite{Goebel2012}.

\section{Preliminaries}
We introduce the following function to facilitate a unified representation of multiple event-triggering conditions.
\begin{definition}\label{def:odot}
	Given $\Xmc\subset\reals[n]$ and $f_1,f_2:\Xmc\to\reals$, define $f_1\otimes f_2:\Xmc\to\reals$ by
	\begin{equation}
		(f_1\otimes f_2)(x)\ceq\left\lbrace \begin{aligned}
			-f_1(x)f_2(x)&\quad\text{if }f_i(x)<0\,\,\forall i\in\{1,2\}\\
			f_1(x)f_2(x)&\quad\text{if }f_i(x)>0\,\,\exists i\in\{1,2\}\\
			{\min\limits_{i\in\{1,2\}} f_i(x)}&\quad{\text{otherwise}}
		\end{aligned}\right. 
	\end{equation}
	
\end{definition}

\subsection{Hybrid Systems}
Basic definitions regarding hybrid systems are borrowed from~\cite{Bernard2020},~\cite{Ricardo2021}, and~\cite{Goebel2012}.
Specifically, we borrow from~\cite[\S~2.1-2.3]{Goebel2012} the concepts of the data of a hybrid system, hybrid time domains, hybrid arcs, and solutions to a hybrid system; from~\cite[\S~3.2]{Ricardo2021} the definitions of stability/attractivity/asymptotic stability of sets for hybrid systems; and from~\cite[\S~2]{Bernard2020} the following notions.
Given a hybrid arc $\phi$, denote $\dom_t\phi$ the projection of $\dom\phi$ onto the $t$-axis and $\dom_j\phi$ the projection of $\dom\phi$ onto the $j$-axis.
For a fixed $j\in\dom_j\phi\minus\{0\}$, $t_j(\phi)$ is the unique time on the $t$-axis such that $(t_j(\phi),j-1)\in\dom\phi$ and $(t_j(\phi),j)\in\dom\phi$; $\Imc[j](\phi)\ceq\{t\in\Rnneg:(t,j)\in\dom\phi\}$ is the largest interval such that $\Imc[j](\phi)\x\{j\}\subset\dom\phi$; and $\Tmc(\phi)\ceq\{t_j(\phi):j\in\dom_j\phi\}$ the set of jump times.
For a fixed $t\in\dom_t\phi$, $\Jmc[t](\phi)\ceq\{j\in\naturals:t_j(\phi)=t\}$ is the set of jump counters of the jumps occurring at time $t$ and hence $\#(\Jmc[t](\phi))$ is the total number of jumps occurring at time $t$.

{A solution $\phi$ to a hybrid system $\Hmc$ is:
\emph{maximal} if there does not exist another solution $\psi$ to $\Hmc$ such that $\dom\phi$ is a proper subset of $\dom\psi$ and $\phi(t,j)=\psi(t,j)$ for each $(t,j)\in\dom\phi$; \emph{complete} if $\sup\dom_t\phi+\sup\dom_j\phi=+\infty$; \emph{precompact} if it is both complete and bounded; \emph{Zeno} if $\sup\dom_j\phi=+\infty$ but $\sup\dom_t\phi<+\infty$.
The set of all maximal solutions to $\Hmc$ is denoted by $\Smc[\Hmc]$.}

Given a hybrid system $\Hmc\ceq(\Cmc,\Fmc,\Dmc,\Gmc)$ and a function $V:\reals[n]\to\Rnneg$ that is locally Lipschitz on a neighborhood of $\Cmc$, we say the growth of $V$ along flows of $\Hmc$ is bounded by $u_{\Cmc}:\reals[n]\to\reals\cup\{-\infty\}$ if $V^\circ(\xi;f)\leq u_{\Cmc}(\xi)$\footnote{$V^\circ(\xi;f)$ denotes the generalized directional derivative of Clarke of a locally Lipschitz function $ V $ at $\xi $ in the direction $ f $, see~\cite{Clarke1990}.} for each $\xi\in\Cmc$ and each $f\in \Fmc(\xi)\cap\Tmc[\Cmc](\xi)$,
and we say the growth of $V$ along jumps of $\Hmc$ is bounded by $u_{\Dmc}:\reals[n]\to\reals\cup\{-\infty\}$ if $V(g)-V(\xi)\leq u_{\Dmc}(\xi)$ for each $\xi\in\Dmc$ and each $g\in \Gmc(\xi)$.

The facilitate presentation of the concept of dwell-time transmission in this letter, we introduce below two useful notions, inspired by~\cite[\S~2.4]{Goebel2012} and~\cite[Definition~2.1]{Bernard2022}.
\begin{definition}[Dwell-Time]\label{def:weak_dwell}
	A hybrid arc $\phi$ is said to have
	\begin{itemize}[leftmargin=*]
		\item a dwell-time if there exists $\tau>0$ such that $|\Imc[j](\phi)|\geq\tau$ for each $j\in\dom_j\phi\minus\{0\}$;
		\item a weak dwell-time if there exists $\tau>0$ such that for each $j\in\dom_j\phi\minus\{0\}$ there exists $k\geq j$ for which $|\Imc[k](\phi)|\geq\tau$.
	\end{itemize}
\end{definition}
\begin{remark}
	Unlike the notions in~\cite[Definition~2.1]{Bernard2022} that ask for uniformity with respect to the initial state, ours are solution-dependent, like~\cite[Proposition~2.34]{Ricardo2021}.
	{We also note that the notion of (weak) dwell-time is stronger than that of Zeno, by virtue of~\cite[Corollary~1]{Zhu2021}.}
\end{remark}

We introduce the following definition to describe multiple jumps occurring at the same time on the $t$-axis, inspired by~\cite[Definition~1]{Biemond2013} and~\cite[Definition~2]{Zhu2021}.
\begin{definition}[{Recursive Jump Maps}]\label{def:G^k}
	Given $\Gmc:\reals[n]\tto\reals[n]$ and $\Dmc\subset\reals[n]$, define $\Gmc^0\ceq\id$, $\Gmc^1\ceq\Gmc$, and $\Gmc^k:\reals[n]\tto\reals[n]$  by $\Gmc^{k+1}(v)\ceq\Gmc(\Gmc^k(v)\cap\Dmc)$ for each $v\in\reals[n]$ and each $k\in\naturals\minus\{0\}$.
\end{definition}
\begin{remark}
	Unlike the notion in~\cite[Definition~1]{Biemond2013} which enforces $\dom\Gmc\subset\Dmc$, ours applies to arbitrary set-valued mappings.
	Unlike the notion in~\cite[Definition~2]{Zhu2021} which is defined for each $k\in\naturals\minus\{0\}$, ours is also defined for $k=0$.
\end{remark}
It turns out that the relaxed geometric-based condition, mentioned in Subsection~\ref{subsec:chaApp}, is equivalent to having a weak dwell-time, as shown next.
\begin{proposition}\label{prop:weak_dwell}
	If a hybrid system $\Hmc\ceq(\Cmc,\Fmc,\Dmc,\Gmc)$ satisfies the hybrid basic conditions (cf.~\cite[Assumption~6.5]{Goebel2012}) and each $\phi\in\Smc[\Hmc]$ is precompact, then each $\phi\in\Smc[\Hmc]$ has a weak dwell-time if and only if there exists $k\in\naturals$ such that $\Gmc^k(\Dmc)\cap\Dmc=\emptyset$.
	$\qed$
\end{proposition}
\begin{proof}
	See Appendix~\ref{app:weak_dwell}.
\end{proof}

\subsection{A Class of Hybrid Systems with Joint Jumping Dynamics}
In this letter, we focus on the class of hybrid systems $\Hmc\ceq(\Cmc,\Fmc,\Dmc,\Gmc)$ taking the following form.
\begin{equation}\label{eqn:mulH}
	\begin{aligned}
		\dot{\xi}&\in \Fmc(\xi)& \xi&\in \Cmc\\
		\xi\pl&\in \Gmc(\xi)\ceq\Gmc[1](\xi)\cup\Gmc[2](\xi) & \xi&\in \Dmc\ceq\Dmc[1]\cup\Dmc[2]
	\end{aligned}
\end{equation}	
where $\Cmc\subset\reals[n]$ is closed relative to $\reals[n]$, $\Fmc:\reals[n]\tto\reals[n]$ is both outer semicontinuous (cf.~\cite[Definition~5.9]{Goebel2012}) and locally bounded (cf.~\cite[Definition~5.14]{Goebel2012}) relative to $\Cmc=\dom\Fmc$, $\Fmc$ is convex-valued on $\Cmc$, $\Dmc[i]\subset\reals[n]$ is closed relative to $\reals[n]$, $\Gmc[i]:\reals[n]\tto\reals[n]$ is both outer semicontinuous and locally bounded relative to $\Dmc[i]=\dom\Gmc[i]$ for each $i\in\{1,2\}$, and
\begin{assumption}\label{ass:NoOverlap}
	$\Gmc[1](\xi)\cap\Gmc[2](\xi)=\emptyset$ for each $\xi\in\Dmc[1]\cap\Dmc[2]$.$\,\,\qed$
\end{assumption}
For each $\phi\in\Smc[\Hmc]$, we define
\begin{equation}\label{eqn:E_Di}
\begin{aligned}
	\Emc[{\Dmc[i]}](\phi)\ceq\{(t,j)\in\dom\phi:\phi(t,j)\in\Dmc[i],\phi(t,j+1)\in\Gmc[i](\phi(t,j))\}
\end{aligned}
\end{equation}
for each $i\in\{1,2\}$ to be the set of hybrid times when the solution jumps according to $\Gmc[i]$.
\begin{proposition}\label{prop:NoOverlap}
	If Assumption~\ref{ass:NoOverlap} holds for $\Hmc$ in~\eqref{eqn:mulH}, then each $\phi\in\Smc[\Hmc]$ satisfies $\Emc[{\Dmc[1]}](\phi)\cap\Emc[{\Dmc[2]}](\phi)=\emptyset$. $\qed$
\end{proposition}
\begin{proof}
	Suppose for sake of contradiction that $\Emc[{\Dmc[1]}](\phi)\cap\Emc[{\Dmc[2]}](\phi)\neq\emptyset$, then there exists $(t,j)\in\dom\phi$ such that $\phi(t,j)\in\Dmc[i]$ and $\phi(t,j+1)\in\Gmc[i](\phi(t,j))$ for each $i\in\{1,2\}$. This implies $\phi(t,j)\in\Dmc[1]\cap\Dmc[2]$ and $\phi(t,j+1)\in\Gmc[1](\phi(t,j))\cap\Gmc[2](\phi(t,j))$, which contradicts with Assumption~\ref{ass:NoOverlap}.
\end{proof}
The above result essentially says that {\eqref{eqn:E_Di}} is well-defined, namely it separates the hybrid times based on the type of jumps a solution undergoes.

Now, suppose we are interested in finding a positive lower bound of the elapsed times between jumps according to $\Gmc[1]$.
Given $\phi\in\Smc[\Hmc]$, we define
\begin{equation}\label{eqn:E_D1}
	\begin{aligned}
		\Delta t(\phi)\ceq\inf\{t_2-t_1:(t_1,j_1),(t_2,j_2)\in\Emc[{\Dmc[1]}](\phi) \text{ for some } 0<j_1<j_2\}
	\end{aligned}
\end{equation}
to denote the greatest lower bound of the elapsed times on the $t$-axis between jumps according to $\Gmc[1]$, excluding an initial jump such that $(0,0)\in\Emc[{\Dmc[1]}](\phi)$ and $\phi(0,1)\in\Gmc[1](\phi(0,0))$. Again, {\eqref{eqn:E_D1}} is well-defined in the sense that its value depends solely on $\Gmc[1]$, by virtue of Proposition~\ref{prop:NoOverlap}.
It is attempting to say that $\Delta t(\phi)>0$ is equivalent to dwell-time.
However, $\Emc[{\Dmc[1]}](\phi)$ is not necessarily a hybrid time domain, hence cannot be used to conclude dwell-time for hybrid arcs (cf.~Definition~\ref{def:weak_dwell}).
To deal with this subtlety, we introduce a deparametrization of a solution as follows.
\begin{definition}[Deparametrization]\label{def:depara}
	Given a solution $\phi$ to $\Hmc$ in~\eqref{eqn:mulH}, a hybrid arc $\psi$ with $\dom\psi\ceq\cup_{(t,j)\in\dom\phi}\Imc[j](\phi)\x\{\#(\Emc[{\Dmc[1]}](\phi)\cap([0,t]\x\{-1,0,\dots,j-1\}))\}$
	is called a $j$-deparametrization of $\phi$. 
\end{definition}
\begin{remark}
	Unlike the notion in~\cite[Definition~3]{Bernard2020} that imposes a same value of $\psi$ and $\phi$ at each $t\in\dom_t\psi$, ours only cares about the underlying hybrid time domains.
\end{remark}

Based on the definition above, we state a result for a deparametrized hybrid arc of a solution to have a dwell-time.

\begin{proposition}\label{prop:dwell}
	Suppose Assumption~\ref{ass:NoOverlap} holds for $\Hmc$ in~\eqref{eqn:mulH} and each $\phi\in\Smc[\Hmc]$ is precompact. 
	For each $\phi\in\Smc[\Hmc]$, if each of the following conditions holds:
	\begin{enumerate}[label=C\arabic*),ref=C\arabic*)]
		\item \label{C1} $\Gmc^k(\Dmc)\cap\Dmc=\emptyset$ for some $k\in\naturals$;
		\item \label{C2} $\Gmc^i(\Dmc[1])\cap\Dmc[1]=\emptyset$ for each $i\in\{1,2,\dots,k-1\}$;
		\item \label{C3} $\Emc[{\Dmc[i]}](\phi)$ is finite for some $i\in\{1,2\}$
	\end{enumerate}
	then each $j$-deparametrization of $\phi$
	has a dwell-time. $\qed$
\end{proposition}
\begin{proof}
	See Appendix~\ref{app:dwell}.
\end{proof}

\subsection{Plant}
We consider a nonlinear continuous-time plant defined by
\begin{equation}\label{eqn:plant}
	\dot{x}_p=f_p(x_p,u)
\end{equation}
where $x_p\in\Xmc[p]\subset\reals[n_p]$ is the plant state, $u\in\Umc\subset\reals[n_u]$ is the input, and the function $f_p:\Xmc[p]\x\Umc\to\reals[n_p]$ is {locally Lipschitz}.

\subsection{Synergistic Controllers}
Suppose we are given a stabilizing synergistic controller for the plant in~\eqref{eqn:plant}.
Following~\cite[\S~7.2]{Ricardo2021}, a synergistic controller $\Hmc[s]\ceq(\Cmc[s],\Fmc[s],\Dmc[s],\Gmc[s],\kappa[s])$ takes the form:
\begin{equation}\label{eqn:Hs}
	\Hmc[s]:\,\left\lbrace\begin{aligned}
		(x_p,q)&\in\Cmc[s]&\dot{q}&\in\Fmc[s](x_p,q)\\
		(x_p,q)&\in\Dmc[s]&q\pl&\in\Gmc[s](x_p,q)\\
		&&u&=\kappa[s](x_p,q)\\
	\end{aligned} \right. 
\end{equation}
where $q\in\Qmc\ceq\{0,1,\dots,\bar{q}\}$ with $\bar{q}\in\naturals\minus\{0\}$.
Letting $\Hmc[0]\ceq(\Cmc[s],\Fmc[0],\Dmc[s],\Gmc[0])$ defined by
\begin{equation}\label{eqn:H0}
	\begin{aligned}
	 	&\pmtx{\dot{x}_p\\\dot{q}}\in \Fmc[0](x_p,q)\ceq\pmtx{f_p(x_p,\kappa[s](x_p,q))\\\Fmc[s](x_p,q)} & (x_p,q)&\in \Cmc[s]\\
		&(x_p\pl,q\pl)\in \Gmc[0](x_p,q)\ceq (x_p,\Gmc[s](x_p,q)) & (x_p,q)&\in \Dmc[s]
	\end{aligned}
\end{equation}
which is the feedback interconnection of~\eqref{eqn:plant} and~\eqref{eqn:Hs}, {we provide below a set of conditions that define a stabilizing synergistic controller:}
	\begin{enumerate}[label=S\arabic*),ref=S\arabic*)]
	\item \label{S1} $\Xmc[p]$ and $\Umc$ are nonempty sets that are closed relative to $\reals[n_p]$ and $\reals[n_u]$, respectively;
	\item \label{S2} {given a compact set $\Amc[0]\subset\Xmc[p]\x\Qmc$}, there exists a continuous function $ V_0:\Xmc[p]\x\Qmc\to\Rnneg$, continuously differentiable on a neighborhood of $\Cmc[s]$, $V_0\inv(0)=\Amc[0]$, $V_0\inv([0,c])$ is compact for each $c\in\Rnneg$, and
	the growth of $V_0$ along flows of $\Hmc[0]$ is bounded by some $u_{{\Cmc[s]}}:\Xmc[p]\x\Qmc\to [-\infty,0]$;
	\item \label{S3} the largest weakly invariant subset of $(\dot{x}_p,\dot{q})\in\Fmc[0](x_p,q)$ in $\cl(u_{\Cmc[s]}\inv(0))$, denoted by $\Psi$, satisfies $\inf\mu(\Psi\minus\Amc[0])\ceqinv\bar{\delta}>0$, where $\mu(x_p,q)\ceq V_0(x_p,q)-\min_{p\in\Qmc}V_0(x_p,p)$ for each $(x_p,q)\in\Xmc[p]\x\Qmc$;
	\item \label{S4} the set $\Cmc[s]\ceq\{(x_p,q)\in\Xmc[p]\x\Qmc: \mu(x_p,q)\leq\delta\}$, the mapping $\Fmc[s](x_p,q)\ceq0$ for each $(x_p,q)\in\Cmc[s]$, the set $\Dmc[s]\ceq\{(x_p,q)\in\Xmc[p]\x\Qmc: \mu(x_p,q)\geq\delta\}$, and the mapping $\Gmc[s](x_p,q)\ceq\{p\in\Qmc:\mu(x_p,p)=0\text{ for some }x_p\in\Xmc[p]\}$ for each $(x_p,q)\in\Dmc[s]$, where $\delta\in(0,\bar{\delta})$;
	\item \label{S5} $\kappa[s]:\Xmc[p]\x\Qmc\to\Umc$ is continuous;
	\item \label{S6} $\Fmc[0](x_p,q)\cap\Tmc[{\Cmc[s]}](x_p,q)\neq\emptyset$ for each $(x_p,q)\in\Cmc[s]\minus\Dmc[s]$
\end{enumerate}
The next result shows the stabilizing capability of $\Hmc[c]$.
\begin{proposition}
	If the conditions~\ref{S1}-\ref{S6} hold then $\Amc[0]$ is globally asymptotically stable for $\Hmc[0]$ in~\eqref{eqn:H0}.$\qed$
\end{proposition}
\begin{proof}
	See, for instance,~\cite[Theorem~1]{Casau2024}.
\end{proof}


\subsection{Event-Triggered Controllers}
As discussed before, we are interested in triggering sporadic transmissions on the communication channels. For sake of simplicity, we only implement event-triggered control on the controller-to-actuator channel. Extensions to include the sensor-to-controller channel can be made following~\cite{Chai2020}.
{The presence of the channel leads us to follow an \emph{emulation-like} approach that derives appropriate event-triggered controllers.}
In this pursuit, the plant no longer has real-time access to the value of $u$ but its sampled version $\hat{u}$, which is generated using zero-order hold devices at the actuator node and can be updated to the value of $u$ when some event-triggering condition is satisfied. Extensions to other holding functions can be made following~\cite{Postoyan2015}.
We model the logic above by a hybrid controller $\Hmc[e]\ceq(\Cmc[e],\Fmc[e],\Dmc[e],\Gmc[e])$ defined by
\begin{equation}\label{eqn:He}
	\Hmc[e]:\,\left\lbrace\begin{aligned}
		(x_p,q,\hat{u},\ell)&\in\Cmc[e]&(\dot{\hat{u}},\dot{\ell})&\in\Fmc[e](x_p,q,\hat{u},\ell)\\
		(x_p,q,\hat{u},\ell)&\in\Dmc[e]&(\hat{u}\pl,\ell\pl)&\in\Gmc[e](x_p,q,\hat{u},\ell)\\
	\end{aligned} \right. 
\end{equation}
satisfying each of the following conditions:
\begin{enumerate}[label=E\arabic*),ref=E\arabic*)]
\item \label{E1} $(\hat{u},\ell)\in\hat{\Umc}\x\Lmc$ with $\hat{\Umc}$ and $\Lmc$ nonempty and closed relative to $\reals[n_u]$ and $\reals[n_{\ell}]$, respectively;
\item \label{E2} there exists a continuous function $\gamma:\Xmc[p]\x\Qmc\x\hat{\Umc}\x\Lmc\to\reals$ such that the set $\Cmc[e]\ceq\{(x_p,q,\hat{u},\ell)\in\Xmc[p]\x\Qmc\x\hat{\Umc}\x\Lmc: \gamma(x_p,q,\hat{u},\ell)\leq 0\}$, the mapping $\Fmc[e](x_p,q,\hat{u},\ell)\ceq(0,f_{\ell}(x_p,q,\hat{u},\ell))$ with $f_{\ell}$ continuous on $\Cmc[e]$, the set $\Dmc[e]\ceq\{(x_p,q,\hat{u},\ell)\in\Xmc[p]\x\Qmc\x\hat{\Umc}\x\Lmc: \gamma(x_p,q,\hat{u},\ell)\geq 0\}$, and the mapping $\Gmc[e](x_p,q,\hat{u},\ell)\ceq(\kappa[s](x_p,q),g_{\ell}(x_p,q,\hat{u},\ell))$ with $g_{\ell}$ continuous on $\Dmc[e]$
\end{enumerate}
Note that the introduction of $\ell$ allows the consideration of dynamic event-triggered strategies.
For sake of convenience, we call $\gamma(\cdot)\geq 0$ the event-triggering condition.

\section{Guaranteeing Dwell-Time Transmission}\label{sec:DT}
Based on the discussions above, we are ready to design a suite of event-triggered synergistic controllers that guarantee dwell-time transmission.
Before proceeding, we first define dwell-time transmission.
To this end, let us define the closed-loop system $\Hmc\ceq(\Cmc,\Fmc,\Dmc,\Gmc)$ resulting from an event-triggered implementation of synergistic controllers, namely
\begin{equation}\label{eqn:H}
	\begin{aligned}
		\Fmc(\xi)\ceq\pmtx{f_p(x_p,\hat{u})\\0\\f_{\ell}(\xi)}&&&\forall \xi\in\Cmc\ceq\{\xi\in\Cmc[e]:(x_p,q)\in\Cmc[s]\}\\
		\Gmc(\xi)\ceq\Gmc[1](\xi)\cup\Gmc[2](\xi) &&& \forall\xi\in \Dmc\ceq\Dmc[1]\cup\Dmc[2]
	\end{aligned}
\end{equation}
where $\xi\ceq(x_p,q,\hat{u},\ell)\in\Xmc[p]\x\Qmc\x\hat{\Umc}\x\Lmc\ceqinv\Xi$,
the set $\Dmc[1]\ceq\Dmc[e]$,
the mapping $\Gmc[1](\xi)\ceq(x_p,q,\kappa[s](x_p,q),g_{\ell}(\xi))$ for each $\xi\in\Xi$,
the set $\Dmc[2]\ceq\{\xi\in\Xi:(x_p,q)\in\Dmc[s]\}$,
and the mapping $\Gmc[2](\xi)\ceq(\Gmc[0](x_p,q),\hat{u},\ell)$
for each $\xi\in\Xi$, such that $\Dmc[i]=\dom\Gmc[i]$ for each $i\in\{1,2\}$.
By virtue of~\cite[Lemma~A.33]{Ricardo2021}, $\Hmc$ satisfies the hybrid basic conditions. Moreover, we observe that
\begin{proposition}\label{prop:NoOverlap_new}
	For $\Hmc$ defined in~\eqref{eqn:H}, it holds $\Gmc[1](\xi)\cap\Gmc[2](\xi)=\emptyset$ for each $\xi\in\Dmc[1]\cap\Dmc[2]$. $\qed$
\end{proposition}
\begin{proof}
	Pick an arbitrary $\xi=(x_p,q,\hat{u},\ell)\in\Dmc[1]\cap\Dmc[2]$, an arbitrary $g_1\in\Gmc[1](\xi)$, and an arbitrary $g_2\in\Gmc[2](\xi)$.
	By construction, $g_1=(x_p,q,\kappa[s](x_p,q),g_{\ell}(\xi))$ and $g_2\in(x_p,q^\ast,\hat{u},\ell)$ for some $q^\ast\in\Gmc[s](x_p,q)$. By definition, $\mu(x_p,q^\ast)=0<\delta$ whereas $\mu(x_p,q)\geq\delta$, and hence $q^\ast\neq q$. This implies $g_1\neq g_2$ and therefore $\Gmc[1](\xi)\cap\Gmc[2](\xi)=\emptyset$.
\end{proof}

Now it is eligible to define $\Emc[{\Dmc[1]}](\phi)$ as per~\eqref{eqn:E_Di} for each $\phi\in\Smc[\Hmc]$ with $\Hmc$ defined in~\eqref{eqn:H} and consequently, we can formalize the notion of dwell-time transmission as follows.
\begin{definition}[Dwell-Time Transmission]\label{def:DTT}
	We say $\Hmc$ in~\eqref{eqn:H} has dwell-time transmission if each $j$-deparametrization of $\phi\in\Smc[\Hmc]$ has a dwell-time.
\end{definition}

We proceed to design a suite of event-triggered synergistic controllers for $\Hmc$ to have dwell-time transmission.
\begin{lemma}\label{lem:dwell_ETC_SYN}
	If each $\phi\in\Smc[{\Hmc}]$ is precompact with $\Hmc$ defined in~\eqref{eqn:H} and each of the following conditions holds:
	\begin{enumerate}[label=C\arabic*'),ref=C\arabic*'),leftmargin=*]
		\item \label{C1'} $\Gmc[i](\Dmc[i])\cap\Dmc[i]=\emptyset$ for each $i\in\{1,2\}$;
		\item \label{C2'} $\Gmc[i](\Dmc[i]\minus\Dmc[3-i])\cap\Dmc=\emptyset$ for each $i\in\{1,2\}$;
		\item \label{C3'} there exists a continuous function $V:\Xi\to\Rnneg$, continuously differentiable on a neighborhood of $\Cmc$, the growth of $V$ along flows of $\Hmc$ is bounded by some $u_{\Cmc}:\Xi\to[-\infty,0]$, and the growth of $V$ along jumps of $\Hmc$ is bounded by some $u_{\Dmc}:\Xi\to[-\infty,0]$ satisfying $u_{\Dmc}(\phi(t,j))=-\delta$ for each $(t,j)\in\dom\phi$ such that $\phi(t,j)\in\Dmc[2]$ and $\phi(t,j+1)\in\Gmc[2](\phi(t,j))$
	\end{enumerate}
	then $\Hmc$ has dwell-time transmission. $\qed$
\end{lemma}
\begin{proof}
	{The conditions~\ref{C1'} and~\ref{C2'} implies that the condition~\ref{C1} holds with $k=2$ and $\Gmc(\Dmc[1])\cap\Dmc[1]=\emptyset$, hence the condition~\ref{C2} holds.
	Now pick an arbitrary $\phi\in\Smc[\Hmc]$.
	Suppose for sake of contradiction that $\Emc[{\Dmc[i]}](\phi)$ is not finite for each $i\in\{1,2\}$.
	We have from the condition~\ref{C3'} and~\cite[Equation~(26)]{Chai2020} that $V(\phi(t,j))\leq V(\phi(0,0)) + \sum_{i=0}^{j}\int_{t_i(\phi)}^{t_{j+1}(\phi)} u_{\Cmc}(\phi(s,i))ds + \sum_{i=1}^{j} u_{\Dmc}(\phi(t_i(\phi),i-1)) \leq V(\phi(0,0)) +\sum_{(s,i)\in\Emc[{\Dmc[2]}](\phi),i\leq j} u_{\Dmc}(\phi(s.i))  \leq V(\phi(0,0)) - \delta j$ for each $(t,j)\in\dom\phi$.
	However, $V(\phi(t,j))<0$ for each $(t,j)\in\dom\phi$ such that $j>\delta\inv V(\phi(0,0))$, which contradicts with $\rge V\subset\Rnneg$.
	Thus, $\Emc[{\Dmc[2]}](\phi)$ is finite and hence the condition~\ref{C3'} holds with $i=2$.
	It follows from Proposition~\ref{prop:dwell} that each $j$-deparametrization of $\phi$ has a dwell-time.
	The proof is completed in view of Definition~\ref{def:DTT}.}
\end{proof}

We show that a naive application of some popular {event-triggering} conditions may \emph{not} ensure dwell-time transmission.
\begin{example}[{Fixed Threshold, cf.~\cite[Theorem~5]{Postoyan2015}}]\label{exp1}
	Pick $n_{\ell}=0$ in the condition~\ref{E1} and define
	\begin{equation}
		\gamma(\xi)\ceq \varrho(|\hat{u}-\kappa[s](x_p,q)|)-\bar{\varrho}
	\end{equation}
	where $\varrho\in\Kinf$ and $\bar{\varrho}>0$.
	The condition~\ref{C1'} holds. However, the condition~\ref{C2'} fails to hold for each $\xi\in\Dmc[2]\minus\Dmc[1]$ and each $g\in\Gmc[2](\xi)$ satisfying $\gamma(g)\geq 0$ so that $g\in\Dmc[1]\minus\Dmc[2]$.
	Moreover, it is not straightforward to apply~\cite[Assumption~1]{Postoyan2015} imposed on continuous feedback laws for the condition~\ref{C3'} to hold.
	 $\qed$
\end{example}

\begin{example}[{Lyapunov-Based, cf.~\cite[Theorem~1]{Zhu2021}}]\label{exp2}
	Pick $n_{\ell}=0$ in the condition~\ref{E1} and define
	\begin{equation}
		\gamma(\xi)\ceq \inner{\nabla V_0(x_p,q)}{(f_p(x_p,\hat{u}),0) - \sigma\Fmc[0](x_p,q)}
	\end{equation}
	where $\sigma\in(0,1)$.
	The condition~\ref{C3'} holds with $V$ in~\cite[Eqn.~(23)]{Zhu2021}.
	However, both the conditions~\ref{C1'} and~\ref{C2'} fail for each $\xi\in\Dmc[1]\minus\Dmc[2]$ and each $g\in\Gmc[1](\xi)$ satisfying $\gamma(g)\geq 0$ so that $g\in\Dmc[1]\minus\Dmc[2]$. $\qed$
\end{example}

\begin{example}[{Dynamic, cf.~\cite[Theorem~3]{Postoyan2015}}]\label{exp3}
	Pick $n_{\ell}=1$ in the condition~\ref{E1} and define
	\begin{equation}
		\gamma(\xi)\ceq \lbar{\ell}-\ell
	\end{equation}
	where $\lbar{\ell}>0$.
	Both the conditions~\ref{C1'} and~\ref{C2'} hold.
	However, it is not straightforward to apply~\cite[Assumption~3]{Postoyan2015} imposed on continuous feedback laws for the condition~\ref{C3'} to hold.
	$\qed$
\end{example}

\section{{Guaranteeing Global Stability}}
Based on the idea of spatial regularization as stated in~\cite[Proposition~2]{Postoyan2015}, we proceed to endow the event-triggered synergistic controllers in Section~\ref{sec:DT} with global stability guarantee.
In view of Examples~\ref{exp1}-\ref{exp3}, we note that it is more convenient to modify the event-triggering condition than to seek an appropriate Lyapunov function candidate. Therefore, our following construction takes advantage of Example~\ref{exp2}, where there exists a Lyapunov function candidate.
Specifically, we combine the Lyapunov-based event-triggering condition in Example~\ref{exp2} with a modified spatial regularization technique.
In this pursuit, we recall Definition~\ref{def:odot} and Example~\ref{exp2} to define $\gamma[1],\gamma[2],\gamma:\Xi\to\reals$ by
\begin{equation}\label{eqn:ETC}
	\begin{aligned}
		\gamma[1](\xi)&\ceq\inner{\nabla V_1(\xi)}{\Fmc(\xi)-\sigma(\Fmc[0](x_p,q),0)}\\
		\gamma[2](\xi)&\ceq |\pi(x_p)|-c\\
		\gamma&\ceq\gamma[1]\otimes\gamma[2]
	\end{aligned}
\end{equation}
where $\xi\ceq(x_p,q,\hat{u})\in\Xi$, $V_1:\Xi\to\reals$ is an extension of $V_0$ such that $V_1(\xi)\ceq V_0(x_p,q)$ for each $\xi\in\Xi$, $\sigma\in(0,1)$, $c>0$, and the function $\pi:\Xmc[p]\to\reals[n_p']$ for some $n_p'\in\naturals$ satisfies each of the following conditions:
\begin{enumerate}[label=$\Pi$\arabic*),ref=$\Pi$\arabic*),leftmargin=*]
	\item \label{Pi1} $\exists c'>0,\,\sup \{V_1(\xi):|\pi(x_p)|\leq c,(x_p,q)\in\Cmc[s]\}\leq c'$;
	\item \label{Pi2} $\sup \{u_{{\Cmc[s]}}(x_p,q):|\pi(x_p)|\geq c,(x_p,q)\in\Xmc[p]\x\Qmc\}<0$
\end{enumerate}

The introduction of $\pi$ modifies the commonly used spatial regularization technique in~\cite[Proposition~2]{Postoyan2015} in the sense that the latter is imposed on the whole state while ours is not, namely the state $q$ is not an argument of $\pi$.
In fact, dwell-time transmission may not be guaranteed without this modification, see~\cite[Corollary~1]{Zhu2021}.
The essence of the modification lies in disabling a second jump when the state is in the jump set of the event-triggered controller.
The main result of this letter is now summarized below.
\begin{theorem}\label{thm}
	Suppose the conditions~\ref{S1}-\ref{S6},~\ref{E1}-\ref{E2}, and~\ref{Pi1}-\ref{Pi2} hold.
	If, for each $\xi\in\Cmc\minus\Dmc$ there exists a neighborhood $\Umc$ of $\xi$ such that $\Fmc(\zeta)\cap\Tmc[\Cmc](\zeta)\neq\emptyset$ for each $\zeta\in\Umc\cap\Cmc$, then the set $\Amc\ceq\{\xi\in\Xi:V_1(\xi)\leq c'\}$ is globally asymptotically stable for $\Hmc$ defined in~\eqref{eqn:H} and $\Hmc$ has dwell-time transmission. $\qed$
\end{theorem}
\begin{proof}
	By construction, $\Hmc$ satisfies the hybrid basic conditions.
	Pick an arbitrary $\phi\in\Smc[\Hmc]$.
	Completeness of $\phi$ is proved using~\cite[Proposition~2.34]{Ricardo2021} by noting: 1) $\Tmc[\Cmc](\xi)\cap\Fmc(\xi)\neq\emptyset$ on $\Cmc\minus\Dmc$; 2) $\Gmc(\Dmc)\subset\Cmc\cup\Dmc$; 3) and $\Fmc$ is locally Lipschitz on $\Cmc$.
	Stability of $\Amc$ is proved by considering the Lyapunov function candidate $V:\Xi\to\reals$ defined by $\xi\mapsto\max\{0,V_1(\xi)-c'\}$.
	Direct manipulation shows that: $u_{\Cmc}:\Xi\to[-\infty,0]$ defined by $u_{\Cmc}(\xi)\ceq 0$ on $\Cmc\cap\Amc$, $u_{\Cmc}(\xi)\ceq (1-\sigma)u_{{\Cmc[s]}}(x_p,q)$ on $\Cmc\minus\Amc$, $u_{\Cmc}(\xi)\ceq-\infty$ on $\Xi\minus\Cmc$; and $u_{\Dmc}:\Xi\to[-\infty,0]$ defined by $u_{\Dmc}(\xi)\ceq 0$ on $\Dmc[1]$, $u_{\Dmc}(\xi)\ceq -V(\xi)$ on $(\Dmc[2]\minus\Dmc[1])\cap V\inv([0,\mu(x_p,q)])$, and $u_{\Dmc}(\xi)\ceq -\mu(x_p,q)$ on $(\Dmc[2]\minus\Dmc[1])\cap V\inv((\mu(x_p,q),+\infty])$.
	It follows from the item (d) of~\cite[Theorem~1]{Chai2020} that $\Amc$ is stable for $\Hmc$.
	Boundedness of $\phi$ is proved by noting that $\hat{u}(\Emc[\Dmc](\phi))\subset\{\hat{u}(0,0)\}\cup\kappa[s](\Umc[0])$ is bounded, where $\Emc[\Dmc](\phi)\ceq\Emc[{\Dmc[1]}](\phi)\cup\Emc[{\Dmc[2]}](\phi)$, $\phi(0,0)\ceq(x_p,q,\hat{u})(0,0)$, and $\Umc[0]\ceq\{(x_p,q)\in\Xmc[p]\x\Qmc: V_0(x_p,q)\leq V(\phi(0,0))+c'\}$ is compact due to the conditions~\ref{S2} and~\ref{Pi1}.
	In view of~\cite[Lemma~3]{Postoyan2015}, there exist $\lbar{\alpha}_V,\bar{\alpha}_V\in\Kinf$ such that $\lbar{\alpha}_V(\norm[\Amc]{\xi})\leq V(\xi)\leq \bar{\alpha}_V(\norm[\Amc]{\xi})$ on $\Xi$.
	Global attractivity of $\Amc$ is then proved by applying the item (d) of~\cite[Theorem~1]{Chai2020} such that $\phi$ approaches the largest weakly invariant set $\Wmc\subset V\inv(r)\cap(u_{\Cmc}\inv(0)\cup (u_{\Dmc}\inv(0)\cap\Gmc(u_{\Dmc}\inv(0))))$ for some $r\in V(\Xi)$. Since $u_{\Cmc}\inv(0)=\Cmc\cap\Amc$ and $u_{\Dmc}\inv(0)=\Dmc[1]\cup(\Dmc[2]\cap\Amc)$, we have $\Wmc\subset V\inv(r)\cap((\Cmc\cap\Amc)\cup(\Dmc[1]\minus\Dmc[2]))=\Amc\cup\Mmc$ with $\Mmc\ceq V\inv(r)\cap((\Dmc[1]\minus\Dmc[2])\minus\Amc)$. Suppose $r>0$, then $\phi(0,0)\in\Mmc$ implies $\phi(t,1)\notin\Mmc$ for some $t\geq 0$. Hence $r=0$ and $\Wmc\subset\Amc$.
	We conclude that $\Amc$ is globally asymptotically stable for $\Hmc$.
	
	Since $\phi$ is precompact and the condition~\ref{C3'} is satisfied by picking $V=V_1$ therein, $\Hmc$ having dwell-time transmission is proved by fulfillment of the conditions~\ref{C1'}-\ref{C2'} in Lemma~\ref{lem:dwell_ETC_SYN}. For sake of brevity, we only present the proof of the case $i=1$ for the condition~\ref{C1'}. The other cases can be proved in a similar vein.
	Specifically, we have that $\xi\in\Gmc[1](\Dmc[1])\cap\Dmc[1]$ implies the existence of some $\xi'=(x_p',q',\hat{u}')\in\Xi$ such that $\gamma[1](\xi)\leq (1-\sigma)u_{{\Cmc[s]}}(x_p',q')<0$ due to~\ref{Pi2}, which contradicts with $\xi\in\Dmc[1]$.
\end{proof}

\section{Application}
Based on Lemma~\ref{lem:dwell_ETC_SYN}, we design an event-triggered synergistic controller for rigid body attitude stabilization with dwell-time transmission.
Let
\begin{equation}\label{eqn:plant0}
	\begin{aligned}
		\dot{x}_p=f_p(x_p,u)\ceq(0.5E(\mf{q})\omega,J^{-1}(\sk(J\omega)\omega+u)) 
	\end{aligned}
\end{equation}
describe the rotational dynamics of a rigid body, where the plant state $x_p\ceq(\mf{q},\omega)\in\sphere{3}\x\reals[3]\ceqinv\Xmc[p]$, the attitude $\mf{q}\ceq(\mf{n},\mf{e})$, the angular velocity $\omega$, the input torque $u\in\reals[3]\ceqinv\Umc$, the function $E:\sphere{3}\to\reals[4\x 3]$ is defined by $E(\mf{q})\ceq\pmtx{-\mf{e}\tp\\\mf{n}I+\sk(\mf{e})}$ with $\sk:\reals[3]\to\so{3}$ defining the bijection between $\reals[3]$ and $\so{3}\ceq\{M\in\reals[3\x 3]:M=-M\tp\}$ such that $\sk(v)w\ceq v\x w$ for each $ v,w\in\reals[3] $, and $J=J\tp\succ 0$ is the inertia matrix.
For the synergistic controller, we pick
\begin{equation}
	\begin{aligned}
		V_0(x_p,q)\ceq 2k_1(1-\varphi(q)\mf{n}) + 0.5\omega\tp J\omega
	\end{aligned}
\end{equation}
where $k_1>0$, $q\in\Qmc\ceq\{0,1\}$, and $\varphi$ is any continuously differentiable function satisfying $\varphi(q)=2q-1$ for each $q\in\Qmc$. Meanwhile, we pick $0<\delta<4k_1\ceqinv\bar{\delta}$ and define
\begin{equation}
	\begin{aligned}
		\kappa[s](x_p,q)\ceq -k_1\varphi(q)\mf{e}-k_2\omega
	\end{aligned}
\end{equation}
where $k_2>0$.
It is straightforward to verify that the conditions \ref{S1}-\ref{S6} hold with
\begin{equation}\label{eqn:A0}
	\Amc[0]\ceq\{(x_p,q)\in\Xmc[p]\x\Qmc:\mf{n}=\varphi(q),\omega=0\}
\end{equation}
For the event-triggering condition, we recall~\eqref{eqn:ETC} and choose $\pi:\Xmc[p]\to\reals[3]$ defined by
\begin{equation}
	\pi(x_p)\ceq\omega
\end{equation}
It is straightforward to verify that the conditions~\ref{E1}-\ref{E2} hold with $\hat{\Umc}=\reals[3]$ and the conditions~\ref{Pi1}-\ref{Pi2} hold with $c'=2k_1+0.5\delta+0.5\eigmax(J)c^2$.
Let $\Hmc$ denote the event-triggered implementation of the synergistic controller and we have the following result. 
\begin{corollary}\label{coro}
	The set $\Amc\ceq\{\xi\in\Xi:V_1(\xi)\leq c'\}$ is globally asymptotically stable for $\Hmc$ and $\Hmc$ has dwell-time transmission. $\qed$
\end{corollary}
\begin{proof}
	Since $\Cmc\minus\Dmc$ is open relative to $\Xi$, $\Tmc[\Cmc](\xi)=\{v\in\reals[4]:\inner{\mf{q}}{v}=0\}\x\reals[3]\x\{0\}\x\reals[3]$ for each $\xi\in\Cmc\minus\Dmc$, implying that $\Tmc[\Cmc](\xi)\cap\Fmc(\xi)\neq\emptyset$ on $\Cmc\minus\Dmc$.
	Applying Theorem~\ref{thm} concludes the proof.
\end{proof}
\begin{remark}
	For the orientation dynamics, we can globally asymptotically stabilize the set $\{(\mf{q},q)\in\sphere{3}\x\Qmc:q\mf{n}\geq 1-c'\}$, which is a \emph{proper} subset of $\sphere{3}\x\Qmc$ if and only if $c\in( 0,\eigmax(J)^{-0.5}(4k_1-\delta)^{0.5}) $. Otherwise, we do not actually have control over the orientation.
\end{remark}

{We execute numerical simulations to verify the effectiveness of our control strategy, for which the code is available in~\cite{zhuSCLgit}. In particular, we compare ours with the following event-triggered synergistic controllers in the literature, where the event-triggering condition equals:
\begin{itemize}[leftmargin=*]
	\item $\gamma[1]$ in~\ref{eqn:ETC}, see~\cite[Remark~5]{Postoyan2015};
	\item $\gamma$ in~\ref{eqn:ETC} with $|\pi(x_p)|$ replaced by $V_1(\xi)$, see~\cite[Corollary~1]{Zhu2021};
	\item the dynamic one in~\cite[Example~7]{Chai2020}
\end{itemize}
}

\begin{figure}[ht!]
	\centering
	\input{plot_V1.tex}
	\includegraphics[width=0.45\textwidth]{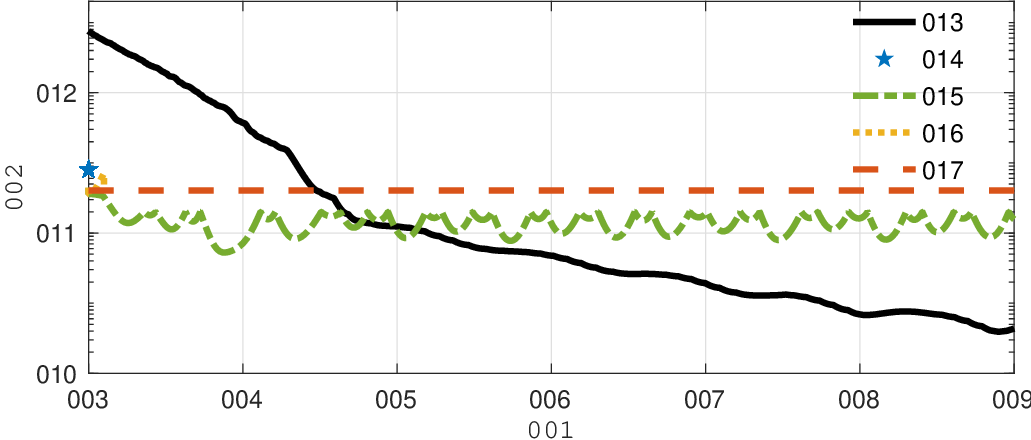}
	\caption{{Comparison of our control strategy with those in~\cite{Postoyan2015,Zhu2021,Chai2020} in terms of the evolution of $V_1$, where $V_1$ being below $c'$ implies $\xi\in\Amc$.\label{fig:plot_V1}}}
\end{figure}

\begin{figure}[ht!]
	\centering
	\input{plot_inter.tex}
	\includegraphics[width=0.45\textwidth]{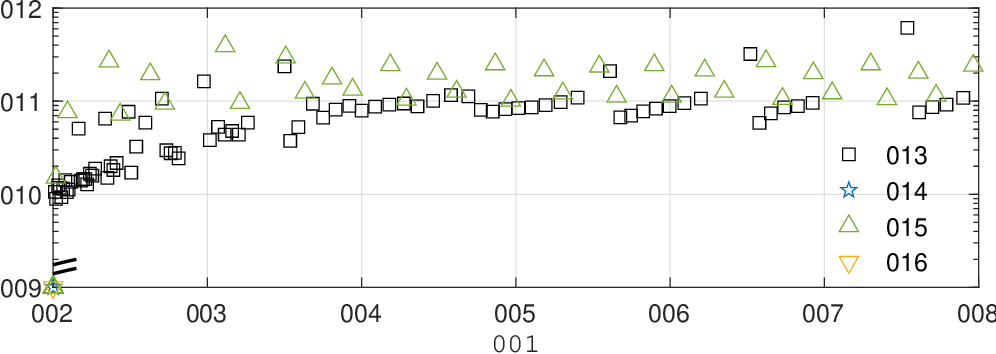}
	\caption{{Comparison of our control strategy with those in~\cite{Postoyan2015,Zhu2021,Chai2020} in terms of the evolution of the elapsed time between transmissions.\label{fig:plot_inter}}}
\end{figure}

{From Figs.~\ref{fig:plot_V1} and~\ref{fig:plot_inter}, we observe that ours ensures completeness of maximal solutions, global asymptotic stability, and dwell-time transmission.
In contrast, the controller in:~\cite[Remark~5]{Postoyan2015} admits a complete discrete solution and hence fails in achieving dwell-time transmission;~\cite[Corollary~1]{Zhu2021} admits a solution that issues \emph{zero} inter-transmission time, hence fails in achieving dwell-time transmission;~\cite[Example~7]{Chai2020} admits a maximal solution that is not complete.
These observations are summarized in Table~\ref{tab:sum}}
\begin{table}[ht!]
	\caption{\label{tab:sum}{Comparison between ours and those in~\cite{Postoyan2015,Zhu2021,Chai2020}.}}
	\centering
	\begin{tabular}{p{140pt}cccc} 
		\hline
		&ours &\cite{Postoyan2015}&\cite{Zhu2021}&\cite{Chai2020}\\
		\hline
		completeness of maximal solutions&\textsurd & \textsurd & \textsurd& \texttimes  \\
		global asymptotic stability&\textsurd &  \texttimes&\textsurd & \texttimes  \\
		dwell-time transmission&\textsurd & \texttimes & \texttimes& \textsurd\\
		\hline
	\end{tabular}
\end{table}

\section{Conclusion}
We synthesize event-triggered control and synergistic control to construct a suite of event-triggered synergistic controllers capable of guaranteeing {both dwell-time transmission and global asymptotic stability}.
The proposed hybrid tools, including geometric-based characterizations of hybrid arcs having a (weak) dwell-time and a deparametrization of hybrid arcs, can be applied to study other combinations of \emph{hybrid} controllers.
One extension of this work is to investigate {other types of event-triggering conditions, besides our proposed combination of Lyapunov-based ones with spatial regularization, that guarantee both dwell-time transmission and global asymptotic stability}.





\appendix
\subsection{Proof of Proposition~\ref{prop:weak_dwell}}\label{app:weak_dwell}

We first prove the ``if'' part. Pick an arbitrary $\phi\in\Smc[\Hmc]$.
Since $\Gmc^k(\Dmc)\cap\Dmc=\emptyset$, $\phi$ cannot be a complete discrete solution. Otherwise, $\phi(0,j)\in\Gmc^j(\Dmc)\cap\Dmc\neq\emptyset$ for each $j\in\naturals$ leads to contradiction.
Since $\Hmc$ satisfies the hybrid basic conditions, it is nominally well-posed due to~\cite[Theorem~6.8]{Goebel2012}.
Together with boundedness of $\phi$, we apply the contrapositive statement of~\cite[Theorem~1]{Casau2022} to arrive at the fact that $\phi$ does not have vanishing time between jumps. This means that either 1) there exists $j\in\naturals$ such that $(t,j)\notin\dom\phi$ for each $t\in\Rnneg$; or 2) there exists $\tau>0$ such that for each $J\in\naturals$ there exists $j\geq J$ for which $t_{j+1}(\phi)-t_j(\phi)\geq\tau$.
For the case 1), completeness of $\phi$ implies that $\phi$ is eventually continuous, namely the last interval of $\dom\phi$ being of the form $\Imc[J](\phi)=[t_J,+\infty)$ with $J=\sup\dom_j<+\infty$. Hence, $t_J<+\infty$ so that $|\Imc[J](\phi)|=+\infty$. This means for each $\tau>0$ and for each $j\in\dom_j \phi\minus\{0\}$, $|\Imc[k](\phi)|\geq\tau$ holds with the selection $k=J$.
For the case 2), we deduce that there exists $\tau>0$ such that for each $j\in\dom_j \phi\minus\{0\}$ there exists $k\geq j$ for which $|\Imc[k](\phi)|=t_{k+1}(\phi)-t_k(\phi)\geq\tau$.
For both cases, $\phi$ has a weak dwell-time, proving the ``if'' part.

We next prove the ``only if'' part by considering the contrapositive statement. Suppose that $\Gmc^k(\Dmc)\cap\Dmc\neq\emptyset$ for each $k\in\naturals$, then we can find $\phi\in\Smc[\Hmc]$ satisfying $\phi(0,j)\in\Gmc^j(\Dmc)\cap\Dmc\neq\emptyset$ for each $j\in\naturals$.
However, for each $\tau>0$ there exists $j=1\in\dom_j \phi\minus\{0\}$ such that $|\Imc[k](\phi)|=0<\tau$ for each $k\geq j=1$. Hence, $\phi$ does not have a weak dwell-time, proving the ``only if'' part.

\subsection{Proof of Proposition~\ref{prop:dwell}}\label{app:dwell}
In view of~\cite[Lemma~A.33]{Ricardo2021}, $\Hmc$ satisfies the hybrid basic conditions.
Note that either $k\in\{0,1\}$ or $\Dmc[1]=\emptyset$ leads to the fact that $\Delta t(\phi)=+\infty$ for each $\phi\in\Smc[\Hmc]$. In the sequel, we focus on the scenario where $k>1$ and $\Dmc[1]\neq\emptyset$.

Picking an arbitrary $\phi\in\Smc[\Hmc]$, we claim that
\begin{equation}\label{eqn:no_E_D1}
	\#(\Emc[{\Dmc[1]}](\phi)\cap(\{t_j(\phi)\}\x\Jmc[{t_j(\phi)}](\phi)))\leq 1
\end{equation}
for each $j\in\dom_j \phi$.
Suppose for sake of contradiction that
there exists $j\in\dom_j \phi$ such that $\#(\Emc[{\Dmc[1]}](\phi)\cap(\{t_j(\phi)\}\x\Jmc[{t_j(\phi)}](\phi)))> 1$. Then there exist $j_1,j_2\in\Jmc[{t_j(\phi)}](\phi)$ with $j_1< j_2$ such that
$	\phi(t_j(\phi),j_i)\in\Dmc[1]$ and $\phi(t_j(\phi),j_i+1)\in\Gmc[1](\phi(t_j(\phi),j_i))$ for each $i\in\{1,2\}$
in view of the definition of $\Emc[{\Dmc[1]}](\phi)$. In view of the condition~\ref{C1}, there are at most $k>1$ jumps at $t_j(\phi)$, implying that there exists $i'\in\{0,1,\dots,k-2\}$ such that $\phi(t_j(\phi),j_2)\in\Gmc^{i'}(\phi(t_j(\phi),j_1+1))$. This implies that $\phi(t_j(\phi),j_2)\in\Gmc^{i'+1}(\phi(t_j(\phi),j_1))\cap\Dmc[1]\subset\Gmc^{i'+1}(\Dmc[1])\cap\Dmc[1]$, which contradicts the condition~\ref{C2} since we find $i=i'+1\in\{1,2,\dots,k-1\}$ such that $\phi(t_j(\phi),j_2)\in\Gmc^i(\Dmc[1])\cap\Dmc[1]\neq\emptyset$.

For an arbitrary $\phi\in\Smc[\Hmc]$,
consider the set
\begin{equation}\label{eqn:no_E_D=1}
	\Tmc[1](\phi)\ceq\{t\in\Rnneg:(t,j)\in\Emc[{\Dmc[1]}](\phi)\,\text{for some}\,j\in\dom_j\phi\}
\end{equation}
which consists of the jump times when $\phi$ jumps exactly once according to $\Gmc[1]$. Then we can rewrite~\eqref{eqn:E_D1} as $\Delta t(\phi)=\inf\{t_2-t_1:t_1,t_2\in\Tmc[1](\phi),0<t_1<t_2\}$.
We assume in the sequel that $\Delta t(\phi)<+\infty$ and proceed with the proof by considering two cases.
Case 1. If the condition~\ref{C3} holds with $\Emc[{\Dmc[1]}](\phi)$ finite, then $\Tmc[1](\phi)$ finite so that $0<\Delta t(\phi)=\min\{t_2-t_1:t_1,t_2\in\Tmc[1](\phi),0<t_1<t_2\}$.
Case 2. If the condition~\ref{C3} holds with $\Emc[{\Dmc[2]}](\phi)$ finite, then there exists $(T,J)\in\dom\phi$ such that $(t,j)\notin\Emc[{\Dmc[2]}](\phi)$ for each $(t,j)\in\dom\phi$ satisfying $t+j\geq T+J$. If $(t,j)\notin\Emc[{\Dmc[1]}](\phi)$ for each $(t,j)\in\dom\phi$ satisfying $t+j\geq T+J$, then we are in Case 1 above.
Otherwise, there exists $(t,j)\in\dom\phi$ satisfying $t+j\geq T+J$ such that $(t,j)\in\Emc[{\Dmc[1]}](\phi)$. Let $T'=t$ and $J'=j+1$.
Consider the hybrid arc $\phi[1]:\Emc[1]\to\reals[n]$ defined by $\phi[1](t,j)\ceq\phi(t,j)$
for each $(t,j)\in\Emc[1]\ceq\dom\phi\cap([0,T']\x\{0,1,\dots,J'\})$, which is a truncated solution of $\phi$ until $(T',J')$. We follow similar arguments in Case 1 to conclude that $\Delta t(\phi[1])>0$.
Consider the hybrid arc $\phi[2]:\Emc[2]\to\reals[n]$ defined by $\phi[2](t,j)\ceq\phi(t+T',j+J')$
for each $(t,j)\in\Emc[2]\ceq\{(t,j)\in\Rnneg\x\naturals:(t+T',j+J')\in\dom\phi\}$, which is a truncated solution of $\phi$ from $(T',J')$ onwards. Since there are no jumps according to $\Gmc[2]$ from $(T',J')$ onwards, $\phi[2]$ can only jump according to $\Gmc[1]$. By virtue of the condition~\ref{C2}, it holds that $\Gmc(\Dmc[1])\cap\Dmc[1]=\emptyset$. It follows from~\cite[Proposition~2.34]{Ricardo2021} that $\Delta t(\phi[2])>0$.
It follows that $\Delta t(\phi)=\min\{\Delta t(\phi[1]),\Delta t(\phi[2])\}>0$.
Therefore, $\Delta t(\phi)>0$ in Case 2.

For an arbitrary $\phi\in\Smc[\Hmc]$, let $\psi$ be an arbitrary $j$-deparametrization of $\phi$. We prove that $\psi$ has a dwell-time.
Note from~\eqref{eqn:no_E_D1} that $|\Imc[j](\psi)|>0$ for each $j\in\dom_j\psi\minus\{0\}$.
Meanwhile, $\Tmc(\psi)=\Tmc[1](\phi)$ for each $\phi\in\Smc[\Hmc]$ with $\Tmc[1]$ defined in~\eqref{eqn:no_E_D=1}.
Hence, pick an arbitrary $\tau\in\reals$ such that $0<\tau<\Delta t(\phi)$ with $\Delta t(\phi)>0$ as proved above, we have that $|\Imc[j](\psi)|\geq \Delta t(\phi)>\tau$ for each $j\in\dom_j\psi\minus\{0\}$. Hence, $\psi$ has a dwell-time.

\section*{Acknowledgment}
This work was supported by the Funda\c{c}\~{a}o para a Ci\^{e}ncia e a Tecnologia (FCT) through FCT Project UIBD/153759/2022.

\bibliographystyle{IEEEtran}
\bibliography{mybib}

\end{document}